\documentstyle[11pt,epsf]{article}                   
\begin{document}
\begin{center}
{\bf Parallelization of a Tree-Code for
the Simulation of Self--Gravitating Astrophysical Systems.}
\end{center}
\centerline{\it R. Capuzzo--Dolcetta$^1$ and P. Miocchi$^{1,2}$}
\centerline{{\small $^{(1)}$Inst. of Astronomy, Univ. ``La Sapienza'',
Rome, Italy.}}
\centerline{{\small $^{(2)}$Dept. of Phys., Univ. ``La Sapienza'', Rome,
Italy.}}
\centerline{{\small E-Mail: dolcetta@astro.uniroma1.it}}
\normalsize
\begin{abstract}
We have preliminary results on the parallelization of a Tree-Code for
evaluating  gravitational forces in N-body astrophysical systems. Our
HPF/CRAFT implementation on a CRAY T3E machine attained an
encouraging speed-up behavior, reaching a value of 75 with 128 processor
elements (PEs).
The speed-up tests regarded the evaluation of the forces among 
$N = 130,369$ particles distributed scaling the actual distribution of a
sample of galaxies.
\end{abstract}

\section{The scientific aim}
In Astrophysics large $N$--body systems are generally {\it self--gravitating},
that is the dynamics of the bodies is strongly influenced by the
gravitational field produced by the bodies themselves.
This `self--influence' makes the exact evaluation of the interactions a
particularly heavy task, since the number of operations needed scales like
$N^2$. To overcome this problem various techniques have been
proposed. Among them, the tree--code algorithm proposed by Barnes \& Hut
(see \cite{BH}, \cite{tesi}) is now widely used in
Astrophysics because it does not require any spatial fixed grid (like, for
example, the `Poisson solver' methods). This makes it suitable to follow very
inhomogeneous and variable (in time) situations,
typical of self-gravitating systems out of equilibrium. 
Furthermore its CPU--time scales like $N\log N$.
\section{The HPF/CRAFT tree--code parallelization}
Tree--codes are difficult to parallelize mainly because gravitation
is a {\it long--range} force and the evaluation of all
the interactions among bodies is not completely separable into a set of
independent tasks (inter-processor communications are inevitable).
Moreover, astrophysical non-uniform distributions imply
great differences in the amount of contributions to the force on each
particle, thus a good {\it load balancing} is hard to be achieved.

In the tree--code we distinguish substantially two parts: i) a
{\it tree--setting} phase where the logical tree structure is built and the
various multipolar coefficients of the cells in which the space is subdivided
are placed in the corresponding locations; ii) a {\it tree--walking} phase
in which the force on each particle is evaluated ``walking'' the tree and
considering all the cells.

Our parallelization was based on a {\it work and
data sharing} approach (using the directives of the HPF/CRAFT language) and
we found that the greatest difficulties in getting good performances are in
the tree--setting phase, in which is not easy to avoid {\it load
unbalancing} among the PEs, due mainly to {\it synchronization} points.
%In order to eliminate such ``bottlenecks''
After various attempts, we adopted a sophisticated scheme whose details can
be found in \cite{3}.

To test the speed-up of our parallelized code on realistic distributions, we
located $N = 130,329$ particles scaling the quite clumped density
distribution of a sample of galaxies in the Northern galactic hemisphere (see
\cite{3}).

\begin{figure}[pt]
  \begin{center}
%%%  \vskip 8.5 true cm
  \leavevmode
  \epsfxsize 9 true cm
  \epsffile{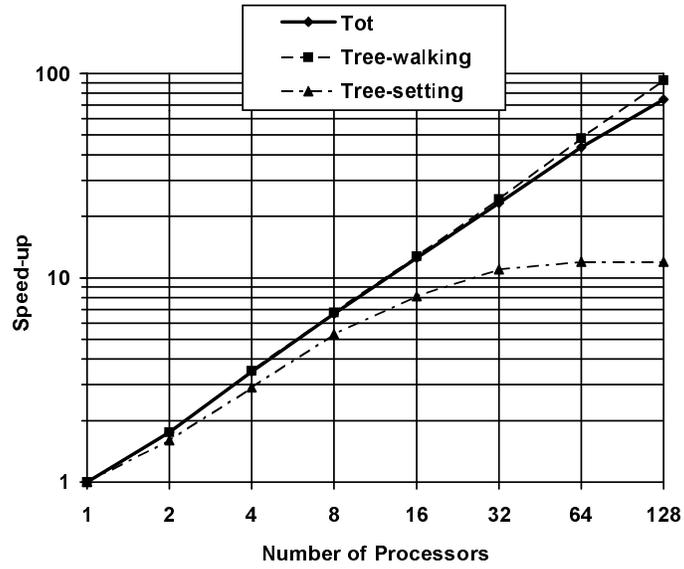}
  \end{center}
  \caption{\protect\small \it 
Measured speed-up for the tree--setting phase, the tree--walking phase and
total.
\label{fig1}}
\end{figure}

\begin{figure}[pt]
  \begin{center}
  \leavevmode
%%%  \vskip 7.5 true cm
  \epsfxsize 8 true cm
  \epsffile{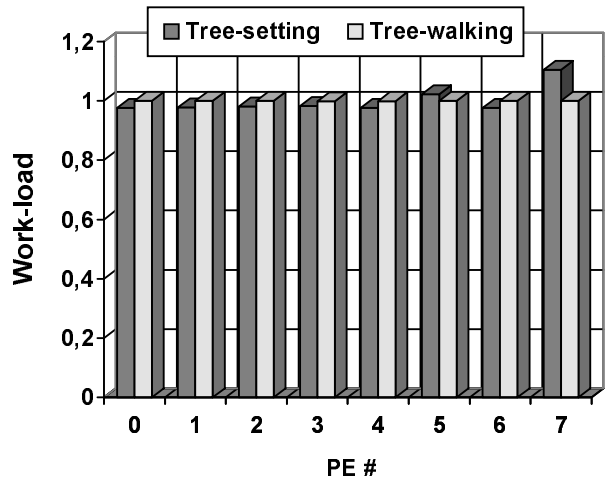}
  \end{center}
  \caption{\protect\small \it 
Normalized work load distribution over 8 processors in both phases.
\label{fig2}}
\end{figure}

In Fig.\ref{fig1} we show the speed-up results obtained on a CRAY
T3E\footnote {This work has been supported by the CINECA--CNAA agreement
(grant {\it cnarm12a}) using the resources at the CINECA
Supercomputing center (Bologna, Italy).} for both phases i) and ii) as well as
the total speed-up.
The tree--setting is confirmed as the most difficult part of the
algorithm to be well parallelized, while the tree--walking speed-up has a
quite good behavior in spite of that it uses intensively remote reading.
In Fig.\ref{fig2} the work load distribution is shown for the run with 8
processors, for both phases. Note how load balancing 
is very good for the tree--walking phase, while for the tree--setting there
are differences among PEs work load which, in any case, do not exceed the 20\%
of the average.

To conclude, the results are rather good: the total speed-up is high enough
and it does not show any flattening, at least for $p\leq 128$.
One has
to consider also that for $p > 16$ the amount of particles per processor is
not that high (less than 5,000). We think that using more particles we would
get even better results.
\noindent
%************
%*  FIG. 1  *
%************
\noindent
%************
%*  FIG. 2  *
%************

\end{document}